\documentclass[ArXiv,AMEGOWP,accept,moreauthors,pdftex,12pt,letterpaper]{Definitions/amego} 

\usepackage{tcolorbox}

\usepackage{setspace}
\usepackage{amsmath}
\usepackage{amssymb}
\firstpage{1} 
\pubvolume{xx}
\issuenum{1}
\articlenumber{5}
\pubyear{2018}
\copyrightyear{2018}
\Title{Prospects for AGN Studies at Hard X-ray through MeV Energies}

\Author{Eileen Meyer$^{1,2,*}$, Justin Finke $^{3,*}$, George Younes $^{4,*}$, Filippo D'Ammando $^{5,*}$, Bindu Rani$^{6,*}$, Sara Buson$^{2,7,*}$, Zorowar Wadiasingh$^{6}$, Ivan Agudo$^{8}$, Volker Beckmann$^{9}$, and Francesco Longo$^{10}$}

\AuthorNames{Firstname Lastname, Firstname Lastname and Firstname Lastname}

\address{$^{1}$ \quad \textbf{Lead Author} |  meyer@umbc.edu; (410) 455-2534\\
$^{2}$ \quad Dept. of Physics, University of Maryland Baltimore County, 1000 Hilltop Circle, Baltimore, MD 21250, USA \\
$^{3}$ \quad U.S. Naval Research Laboratory,Washington, DC, USA\\
$^{4}$ \quad Department of Physics, The George Washington University, Washington, DC 20052, USA\\
$^{5}$ \quad INAF - IRA, Bologna, 40129, Italy\\
$^{6}$ \quad NASA Goddard Space Flight Center, Greenbelt, MD, 20771, USA \\
$^{7}$ \quad University of W\"{u}rzburg, 97074 W\"{u}rzburg, Germany \\
$^{8}$ \quad Instituto de Astrof\'{\i}sica de Andaluc\'{\i}a (CSIC),  
Apartado 3004, E--18080 Granada, Spain \\
$^{9}$ \quad Institut National de Physique Nucléaire et de Physique des Particules (IN2P3)
CNRS, Paris, France\\
$^{10}$ \quad Dipartimento di Fisica, Universit a degli Studi di Trieste, via A. Valerio 2, I-34100, Trieste, Italy \\
$^{11}$ \quad INFN, Sezione di Trieste, via A. Valerio 2, I-34100, Trieste, Italy\\
$^{*}$ \quad \emph{principal authors}
}

\simplesumm{Formation and Evolution of Compact Objects; Galaxy Evolution} 

\abstract{This White Paper explores advances in the study of Active Galaxies which will be enabled by new observing capabilities at MeV energies (hard X-rays to $\gamma$-rays; 0.1$-$1000 MeV), with a focus on multi-wavelength synergies. This spectral window, covering four decades in energy, is one of the last frontiers for which we lack sensitive observations. Only the COMPTEL mission, which flew in the 1990s, has significantly probed this energy range, detecting a handful of AGN. In comparison, the currently active \emph{Fermi} Gamma-ray Space Telescope, observing at the adjacent range of 0.1$-$100 GeV, is 100-1000 times more sensitive.  This White Paper describes advances to be made in the study of sources as diverse as tidal disruption events, jetted AGN of all classes (blazars, compact steep-spectrum sources, radio galaxies and relics) as well as radio-quiet AGN, most of which would be detected for the first time in this energy regime. New and existing technologies will enable MeV observations at least 50-100 times more sensitive than COMPTEL, revealing new source populations and addressing several open questions, including the nature of the corona emission in non-jetted AGN, the precise level of the optical extragalactic background light, the accretion mode in low-luminosity AGN, and the structure and particle content of extragalactic jets. }

\begin{document}
\setstretch{1.0}

\clearpage

\section{Introduction}

\vspace{-5pt}
\noindent
Active Galactic Nuclei (AGN) are the small fraction of galaxies with an actively accreting black hole at their center \citep{netzer2015}. While high accretion rate systems can lead to extreme luminosities (up to 10$^{46}$ erg/s), accretion rates vary from a maximum near or at the Eddington limit ($L_{Edd}$) down to very low levels (10$^{-6}L_{Edd}$). 
In perhaps 10\% of AGN,  bipolar jets of ionized plasma are ejected with highly relativistic speeds very near the black hole and extending out hundreds to thousands of parsecs, often well beyond the host galaxy \citep{urry95}. The latter are termed `radio-loud' AGN due to the radio synchrotron emission produced by the jets; the jet spectrum extends from radio up to X-ray/gamma-ray energies and tends to dominate the overall spectral energy distribution (SED) due to Doppler boosting of the emission, especially in sources with jets aligned with a small orientation angle to earth (`blazars'). 

This White Paper is concerned with the prospects for AGN studies at MeV energies, or approximately 0.1$-$1000 MeV. It is instructive to note that jetted AGN are the dominant source class at even higher (GeV) energies: approximately 3100 of the ~3600 associated sources in the latest \emph{Fermi}/LAT catalog (at 0.1$-$1000 GeV) have been identified as radio-loud AGN, nearly all blazars. 
At MeV energies, only the COMPTEL mission has conducted a major sky survey. Operating from 0.75$-$30 MeV, the observatory detected some 63 sources from 1991-1995, of which 10 were jetted AGN (none were radio-quiet AGN) \citep{schonfelder2000}. Successor missions currently in development include the All-sky Medium Energy Gamma-ray Observatory (AMEGO), a probe mission concept in the range $\sim 0.1$\ MeV to 10 GeV.  Combining two different $\gamma$-ray detection methods (Compton scattering and pair production), it will have several orders of magnitude better sensitivity than COMPTEL, as well as unprecedented spectral resolution and polarization sensitivity \citep{amego}.  While AMEGO is not the exclusive focus of this white paper, we assume comparable capabilities for the science cases we present. In the sections below we briefly outline some of the discoveries and investigations in AGN studies which will be made possible by sensitive imaging observations at MeV energies.  We note that most of these science cases are inherently multi-wavelength in nature and will be far more impactful if an MeV mission flies concurrently with observatories like CTA, LSST, and SKA.



\section{Radio Loud AGN}
\vspace{-2pt}
\noindent
\textbf{Blazar Population Studies -- }
While the \emph{Fermi}/LAT has revolutionized the study of high-energy emission from jetted AGN by detecting thousands of sources (compared with 51 total from predecessor EGRET), many open questions remain due to the `missing' several orders of magnitude adjacent to the Fermi band stretching from soft X-rays to 0.1 GeV, which includes at least half of the inverse Compton spectrum emitted by these objects.  Indeed, there is evidence that the most powerful jetted AGN have the peak of their $\gamma$-ray emission in the MeV band \citep{ajello2016}, with {\em Fermi} detecting only the 'tail' or not detecting the sources at all.  This has obvious implications for our understanding of this population -- in particular any spectral evolution with redshift, which could tell us about black hole fueling and the changing environment with redshift \citep{inoue2009}. AMEGO, as an example, should detect up to several thousand blazars, providing a complimentary catalog of sources to Fermi \citep{inoue2015}. 

The broad high-energy component (from X-rays to GeV or even TeV energies) observed in jetted AGN is generally attributed to inverse Compton (IC) upscattering of a population of lower-energy 'seed' photons by relativistic electrons in the jet; these seed photons are not clearly identified in all cases but could be synchrotron photons produced by the jet itself (synchrotron self-Compton or SSC) or an external photon field (disk, galaxy, broad-line emitting clouds, molecular gas) leading to external Compton or EC emission. Alternatively, `hadronic' models, in which the jet plasma includes a significant population of relativistic protons, may also explain the gamma-ray emission \citep{aha2000}. MeV observations can constrain hadronic models for the emission from blazars in several ways.  For low-power BL Lac objects, an SED component is expected in the MeV band which is not expected in leptonic models \citep{petropoulou2015}.  Further, since in hadronic models $\gamma$-rays are produced from the synchrotron emission from decay components of p$\gamma$ interactions, the $\gamma$-rays are also expected to be polarized in hadronic models \citep{zhang2013,zhang2014,zhang2015,zhang2016_1,zhang2016_2}.  Thus, detection of variability and polarization by AMEGO could also constrain hadronic models 
(see related White Paper lead by B. Rani).

\vspace{5pt}
\noindent
\textbf{Radio Galaxies/Misaligned Jets -- }
Population studies of jetted AGN will also be aided by MeV observations of \emph{misaligned} jets, usually termed 'radio galaxies' (RG) since the optical emission, being not nearly so beamed, does not outshine the host galaxy. In these sources (unlike blazars) it is generally the case that the jet is well-resolved by instruments like the VLBA, VLA and HST (and future SKA) on scales from tens to hundreds of parsecs. As we are still in ignorance about many of the primary characteristics of jets -- particle content, jet velocity structure, particle acceleration mechanisms -- the opportunity to examine how the high-energy spectrum changes with jet orientation angle \citep{geo2003,meyer2011}, as well as the ability to connect high-energy variability to changes in the jet structure is extremely important. In the latter case, multi-wavelength observations might allow us to localize where in the jet the high-energy emission arises. In M87, for example, a TeV flare was observed during a years-long outburst from optical to X-rays in a knot 100 parsecs from the black hole, suggesting that the very high-energy emission might arise much further downstream that typically thought \citep{harris2009} -- see Figure~1. In the near future simultaneous observations with e.g., CTA, SKA, and AMEGO will enable study of a much larger number of targets to solve this issue. So far, the \emph{Fermi}/LAT has detected only a handful of nearby RG, likely because the high energy spectrum peaks in the MeV regime, as suggested by the soft GeV spectrum. 
Observations in the MeV will be key to greatly increasing the number of RG detected in $\gamma$-rays and to better characterizing the broad band SED of these sources.  In addition, such observations will allow us to determine the contribution of misaligned jets to the extragalactic $\gamma$-ray background in the almost unexplored 0.1$-$100 MeV energy range.

\vspace{5pt}
\noindent
\textbf{Narrow-line Seyfert 1 galaxies --}
The discovery by {\em Fermi}/LAT of variable $\gamma$-ray emission from a few radio-loud narrow-line Seyfert 1 galaxies (NLSy1), a rare type of Seyfert characterized by peculiar optical properties \cite{osterbrock85, pogge00} revealed the presence of a probable third class of AGN with relativistic jets \citep{abdo09, dammando16}. 
The discovery of a relativistic jet in a class of AGN thought to be hosted in spiral galaxies \citep[e.g.][]{deo2006}, with a BH mass typically between 10$^6$--10$^7$ M$_{\odot}$, challenges the current knowledge on how the jets are generated and developed \citep{boettcher2002}. 
However, near-infrared observations of 
two $\gamma$-ray emitting NLSy1 reveal elliptical host galaxies with a BH mass larger than 10$^{8}$ M$_{\odot}$ \cite{damm2017, damm2018}, in agreement with the theoretical models for the production of relativistic jets. A much larger sample of jetted NLsy1 are needed to follow up these studies. The peak of the IC emission in these sources is expected to lie in the MeV regime, as confirmed by the soft GeV-band spectrum in the few that have been detected. MeV band observations will allow us to both detect a larger number of NLSy1 with respect to {\em Fermi}/LAT and better determine the high-energy peak of the SED of these sources and thus to constrain the emission mechanisms for producing their high-energy emission.   

\vspace{5pt}
\noindent
\textbf{Compact Steep Spectrum Objects -- } A fourth class of jetted AGN are the Compact Steep Spectrum or CSS objects.  Thought to be very young radio galaxies (age $<$ few thousand years), one has recently been reported detected by the \emph{Fermi}/LAT \citep{migliori_css}.  As in the previous three cases, the very soft LAT band spectrum (and earlier failure to detect these sources in the Fermi band \citep{dammando16_css}) suggests that these sources have a high-energy emission peak in the MeV. Unlike blazars, very little is known about the emission properties of CSS sources, especially at high energies.


\vspace{5pt}
\noindent
\textbf{Giant Radio Lobes and Relics --} Powerful and/or very nearby radio galaxies often exhibit very large `lobes' of slowed, radio synchrotron-emitting plasma which emit $\gamma$-rays. A striking example is shown in Figure~1: the radio lobes of Centaurus A are approximately 7$^\circ$ in extent and have been detected by the \emph{Fermi}/LAT \citep{fermi_cena}. This would represent one of the few \emph{resolveable} MeV sources with an instrument like AMEGO (expected angular resolution $\sim$3$^\circ$ at 1 MeV) or AdEPT, a pair telescope with superior angular resolution to AMEGO above ~10 MeV \citep{adept}.  These radio lobes emit from X-rays to $\gamma$-rays through inverse-Compton upscattering of ambient photon fields pervading the Universe; in particular the Cosmic Microwave Background (CMB) and the extragalactic background light (EBL), the latter peaking both in the infrared and optical. Upscattered CMB photons dominate from X-rays to MeV $\gamma$-rays, while the EBL (due to the higher energy of the seed photons) dominates above ~1 GeV.

\begin{figure}[!t]
\centering
\includegraphics[width=6.1in]{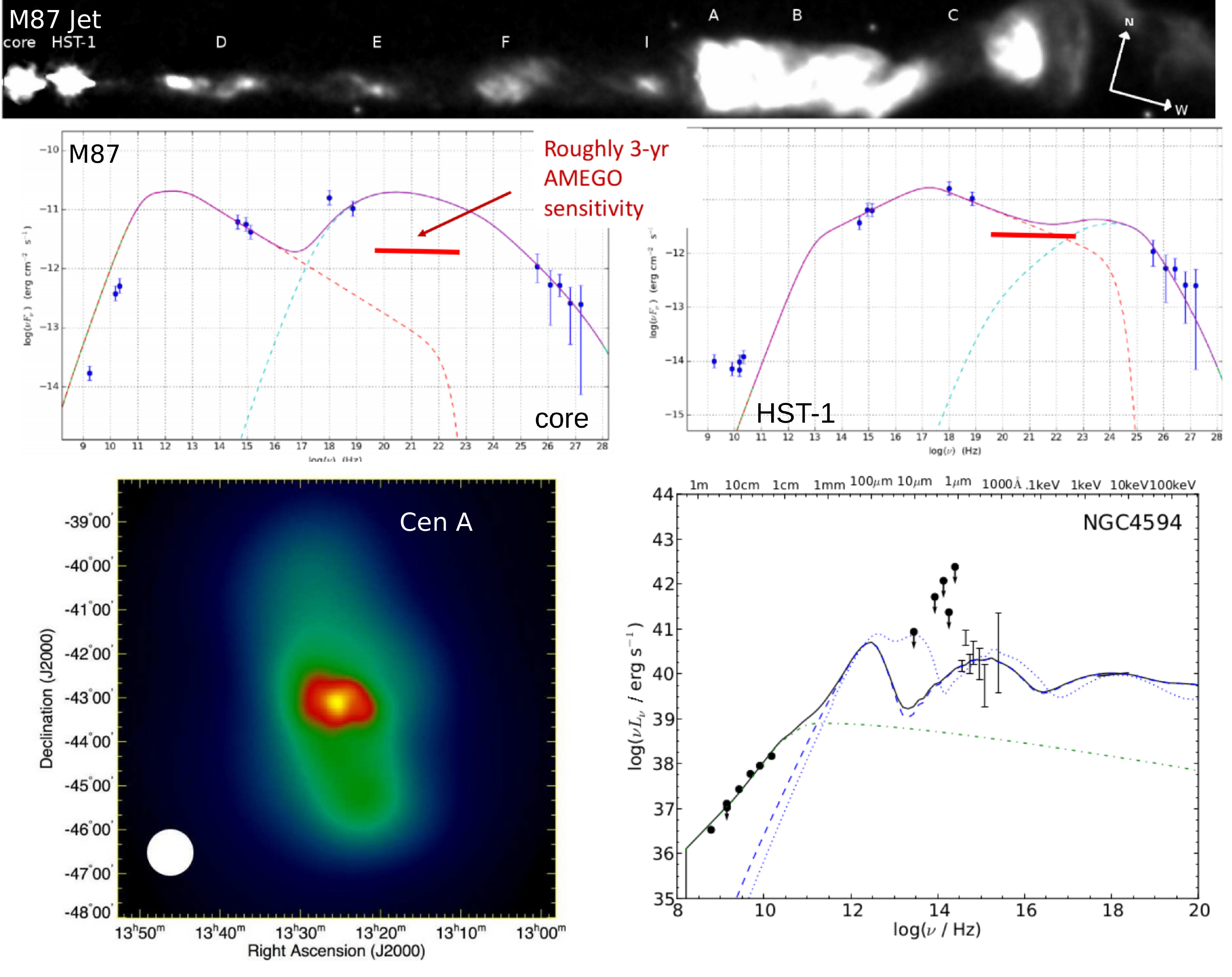}
\caption{\footnotesize \emph{Upper Figures:} An HST image of the jet in nearby galaxy M87 \citep{meyer2013}.  The knot labeled HST-1 is approximately 100 pc (projected) downstream of the 'core' (emission arising very near to the black hole). In the panels below are the SEDs for the core (left) and HST-1 (right) \citep{dejong2015}.  In each, the Fermi observations are the total flux from the source: it is not known where the GeV emission originates.  The model curves show how critical are MeV observations for filling in a very large gap in the spectral coverage which could allow us to differentiate. \emph{Lower Left:} Fermi/LAT image of the resolved radio-emitting lobes of Centaurus A \citep{fermi_cena}.  The GeV emission arises from inverse Compton upsattering of the EBL.  In the MeV band, the same lobes will be visible (and resolvable with AMEGO) due to upscattered CMB photons. \emph{Lower Right:} One possible model for an ADAF in low-luminosity AGN NGC 4594 \citep{nemmen14}.  ADAF models generally predict emission up to MeV energies and beyond; measurement of the level and spectral shape will allow us to constrain these models in a band with no competition from other emission sources.  }
\end{figure}   


 A key feature is that the peak of the IC/CMB lobe emission in most cases is at MeV energies. Extrapolation of the X-ray spectrum of the lobes of radio galaxies \citep{croston2005} to the MeV range indicates that AMEGO will detect a number of these sources (including `relic' sources for which the central AGN no longer fuels an active jet). Measuring the level of the IC/CMB emission is the only way to measure the strength of the magnetic field in the lobes, and to interpret the GeV spectrum to measure the EBL photon density \citep{georgan2008}. Measuring the EBL this way has been attempted with Fornax A \citep{fermi_fornaxA2016}, however in that case it was suspected that the GeV spectrum was dominated by $\gamma$-rays produced by the decay of pions produced from proton-proton interactions.  While the critical measurement of the IC/CMB normalization can and has been done using X-ray observations, MeV is preferable due to the lack of confounding background sources implanted in the giant lobes (which can be arcminute to degrees in scale) and for the simple fact that the emission is peaking in this band.   

\section{Tidal Disruption Events}
\vspace{-5pt}

When a star passes close to a super-massive black hole (SMBH), tidal forces will tear the star apart \cite{fr76}.  For SMBHs with $M_{\mathrm{BH}} \lesssim 10^{8}$\,M$_{\odot}$, the streaming debris will emit a flash of radiation known as a tidal disruption event (TDE; \cite{h75}).  TDEs are unique probes of SMBHs in galaxies that are too distant for resolved kinematic studies and are not currently actively accreting gas \cite{r88}.
The discovery by \textit{Swift} of TDEs associated with newly formed relativistic jets \cite{bkg+11,ltc+11,bgm+11,zbs+11,ckh+12} indicates that these systems can also serve as valuable probes
of accretion physics, including possible ``state transitions’' \cite{zbm+13,pcl+15}.  With its wide field-of-view and MeV sensitivity, AMEGO will be a powerful discovery engine for the non-thermal emission generated in these TDE jets, helping to reveal the conditions necessary for the formation of such relativistic ejecta (e.g., black hole spin, stellar magnetic field).  The detection of 9 such nearby events at hard X-ray energies by the Swift BAT \cite{Hryniewicz+2016} suggests a large discovery rate by AMEGO, which can then be directly compared to the sample of thermal (disk-dominated) events discovered by future wide-field optical (e.g., LSST) and X-ray (e.g., eROSITA) surveys.

\section{Radio Quiet AGN}
\vspace{-2pt}


\noindent
{\bf Corona Emission --} X-ray emission in radio quiet AGN is thought to be dominated by Compton scattering of accretion disk photons by a population of thermal electrons in a corona that sandwiches the disk. The X-ray spectrum is generally a power-law with a cut-off at $\approx 50$-$200$\ keV  for Seyfert galaxies \citep{fabian2015} with the exact energy related to the temperature of the corona.  Some sources observed by hard X-ray instruments OSSE, INTEGRAL, NuSTAR, and BAT \citep[e.g.,][]{johnson1997,vasudevan2013,malizia2014} could be detectable by AMEGO based on a simple extrapolation.  It has been theorized that the temperature of the corona is regulated by $\gamma\gamma$ pair production as a sort of ``thermostat'' \citep[e.g.,][]{svensson1984,zdz1985,stern1995,coppi1999}.  That is, as the temperature and density of the coronal plasma increases, more pairs will be created.  Since more energy goes into pair creation, this limits the increase in temperature.  In this case a component above the cutoff energy in the $\gamma$-rays would be expected, from Compton scattering of disk photons by the newly created pairs.  This component could be detectable by a sensitive MeV observatory, confirming or ruling out this model.

\vspace{3pt}
\noindent
{\bf Low Luminosity AGN --} The dimness of low luminosity AGN (LLAGN) is due to an underfed supermassive BH      \citep[see][for a review]{ho2008}. At low accretion rates, the density in the accretion disk becomes too low for radiative cooling to be effective. The trapped heat will expand the inner parts of the accretion disk into a pressure-supported optically-thin geometrically-thick accretion flow, with low radiative efficiency \citep{yuan14}. The flow would then be either advected onto the BH -- the Advection Dominated Accretion Flow (ADAF) solution, Figure~1 \citep{yuan2004} -- or released as a jet and/or outflow \citep{blandford99MNRAS:jet,narayan00ApJ}.

Inverse Compton scattering of synchrotron photons by hot electrons in the flow is the dominant energy loss in radiatively-inefficient accretion models. The energy of the particles in the flow is predicted to peak in the soft MeV range, implying an exponential cutoff in the photon spectrum around the same energy \citep{yuan14}. Direct measurement of the high energy cutoff would enable a direct estimate of the plasma temperature and density in the flow. Fitting the full multiwavelength SED of LLAGN along with the soft $\gamma$-ray spectrum may finally lead to an estimate of the accretion flow particle density profile in LLAGN, which is crucial in understanding feedback in these systems. Given the low sensitivity of past and current soft $\gamma$-ray instruments, the high energy cutoff has been measured in only one LLAGN so far, NGC~3998; $E_{\rm cutoff}=110$~keV \citep{younes19ApJ}, in addition to a few lower limits $E_{\rm cutoff}> 150$~keV by NuSTAR \citep{ursini15MNRAS,young18MNRAS}. Future high sensitivity soft $\gamma$-ray instruments will enable a detailed look into the depth of the hot accretion flow of LLAGN, revealing the physical properties and the interplay of the inflow-outflow mass governing the appearances of the central engines of these sources.

\externalbibliography{yes}
\bibliographystyle{yahapj}
\bibliography{references}

\end{document}